%% file: main.tex
\documentclass[runningheads,orivec]{llncs}

\usepackage[utf8]{inputenc}
\usepackage[T1]{fontenc}
\usepackage{float}
\usepackage{mysl}
\usepackage{graphicx}
\usepackage[frozencache]{minted}
\usepackage{listings}
\usepackage{xcolor}
\usepackage{multirow}
\usepackage{booktabs}
\usepackage{todonotes}
\usepackage{fontawesome}
\usepackage{xspace}
\usepackage[numbers,sort&compress]{natbib}
\usepackage{array}
\usepackage{circledsteps}
\usepackage{lstautogobble}
\usepackage{wrapfig}
\usepackage{orcidlink}
\usepackage{hyperref} 

\graphicspath{ {./images/} }

\definecolor{bluekeywords}{rgb}{0.13, 0.13, 1}
\definecolor{greencomments}{rgb}{0, 0.5, 0}
\definecolor{redstrings}{rgb}{0.9, 0, 0}
\definecolor{graynumbers}{rgb}{0.5, 0.5, 0.5}
\definecolor{tealgreen}{HTML}{198269}
\definecolor{lightgreen}{RGB}{66, 199, 30}
\definecolor{burgundy}{HTML}{530021}
\definecolor{rosered}{HTML}{D60B47}
\definecolor{mygrey}{HTML}{595959}

\lstdefinelanguage{gil}{
  otherkeywords={*,+,==},
  keywords=[1]{predicate,function},
  keywordstyle=[1]\color{rosered},
  keywords=[2]{list,llen},
  keywordstyle=[2]\color{burgundy},
  keywords=[3]{==,++,len,*,:=,+},
  keywordstyle=[3]\color{tealgreen}
}

\newcommand{\PreserveBackslash}[1]{\let\temp=\\#1\let\\=\temp}
\newcolumntype{C}[1]{>{\PreserveBackslash\centering}p{#1}}
\newcolumntype{R}[1]{>{\PreserveBackslash\raggedleft}p{#1}}
\newcolumntype{L}[1]{>{\PreserveBackslash\raggedright}p{#1}}

\def\releaseMode{0}  
\def\releaseMode{1}  

\def\extendedVersion{0}
\def\extendedVersion{1}
\newcommand\ifExtended[2]{
  \if\extendedVersion1
    #1
  \else
    #2
  \fi
}

\if\releaseMode0
    \makeatletter%
    \@mparswitchfalse%
    \makeatother%
    \normalmarginpar

    \newcommand{\towrite}[2][]{\todo[inline, color=blue!30, #1]{#2}}
    \newcommand{\unsure}[2][]{\todo[color=red!30, #1]{#2}}
    \newcommand{\review}[2][]{\todo[color=green!30, #1]{#2}}
    \newcommand{\sacha}[2][]{\review[#1]{\textit{Sacha:} #2}}
    \newcommand{\pg}[2][]{\review[#1]{\textit{Philippa:} #2}}
    \newcommand{\jules}[2][]{\review[#1]{\textit{Jules:} #2}}
\else
    \renewcommand{\todo}[2][]{}
    \newcommand{\towrite}[2][]{}
    \newcommand{\unsure}[2][]{}
    \newcommand{\review}[2][]{}
    \newcommand{\sacha}[2][]{}
    \newcommand{\pg}[2][]{}
    \newcommand{\jules}[2][]{}
\fi

\def\sectionautorefname{\normalcolor{\S}\kern-0.7ex\color{blue}}
\def\subsectionautorefname{\normalcolor{\S}\kern-0.7ex\color{blue}}

\definecolor{Green}{RGB}{0, 180, 6}

\definecolor{MatchErrFG}{RGB}{255, 0, 0}
\definecolor{MatchErrBG}{RGB}{210, 210, 210}

\begin{document}

\if\releaseMode0
  \tableofcontents
  \newpage

  \listoftodos

  \section*{TODO key}
  \todo[inline, nolist]{To do / task}
  \towrite[nolist]{Placeholder; to be written}
  \unsure[inline, nolist]{Question / uncertainty}
  \review[inline, nolist]{Note from reviewer}


  \newpage
  \
  \newpage
  \setcounter{page}{1}
\fi

\if\extendedVersion0
  \title{Gillian Debugging: Swinging Through the (Compositional Symbolic Execution) Trees}
  \titlerunning{Gillian Debugging}
\else
  \title{Gillian Debugging: Swinging Through the (Compositional Symbolic Execution) Trees, Extended Version}
  \titlerunning{Gillian Debugging (Extended)}
\fi

\author{
  Nat Karmios%
  \,\orcidlink{0009-0000-3582-2483}%
  \email
\and
  Sacha-Élie Ayoun%
  \,\orcidlink{0000-0001-9419-5387}
\and
  Philippa Gardner%
  \,\orcidlink{0000-0002-4187-0585}
}
\authorrunning{Nat Karmios \and Sacha-Élie Ayoun \and Philippa Gardner}

\institute{
  Imperial College London, UK \\
  \email{n.karmios@ic.ac.uk}
}

\maketitle

\vspace*{-3mm}
\begin{abstract}
In recent years, compositional symbolic execution (CSE) \linebreak
tools have been growing in prominence and are becoming more and more applicable to real-world codebases.
Still to this day, however, debugging the output of these tools remains difficult, even for specialist users.
To address this, we introduce a debugging interface for symbolic execution tools, integrated with Visual Studio Code and the Gillian multi-language CSE platform, with strong focus on visualisation, interactivity, and intuitive representation of symbolic execution trees.
We take care in making this interface tool-agnostic, easing its transfer to other symbolic analysis tools in future.
We empirically evaluate our work with a user study, the results of which show the debugger's usefulness in helping early researchers understand the principles of CSE and verify fundamental data structure algorithms in Gillian.
\end{abstract}
\vspace*{-3mm}


\input{sections/01_intro}

\input{sections/02_related-work}
\input{sections/03_background}

\input{sections/04_debug-intro}
\input{sections/05_design}

\input{sections/06_impl}
\input{sections/07_user-study}
\input{sections/08_future}

\input{sections/09_conclusion}
\todo{This needs to fit into 17 pages!}

\vspace*{-0.2cm}
\subsection*{Data Availability Statement}
\vspace*{-0.1cm}
{\footnotesize
An artifact containing Gillian binaries, the debugger VSCode extension, and example WISL programs (including the exercises used in evaluation) is available at \cite{artifact}.
}
\vspace*{-0.3cm}
\subsection*{Acknowledgements}
\vspace*{-0.1cm}
{\footnotesize
We extend a heartfelt thanks to Opale Sj{\"o}stedt, Simon Park and Diego Cupello for their tireless work in creating exercises for the user study, as well as the participants of the study for making our evaluation possible;
Jessica Shi and Alastair Donaldson for their advice on constructing the user study and this paper respectively;
Petar Maksimovi{\'c} for his support in this project's inception;
and Radu Lacraru and Matthew Ho, whose MEng projects formed the engineering foundation that this project was built upon.

This work was supported by funding from Gardner's UKRI fellowship `Verified Trustworthy Software Specification' and grants from Meta and Amazon.
}

\bibliography{references}

\if\extendedVersion1
  \clearpage
  \appendix
  \input{sections/appendix}

\fi

\end{document}

%% file: sections/01_intro.tex

\vspace*{-0.25mm}
\section{Introduction}
\label{sec:intro}
\vspace*{-0.25mm}

Symbolic execution (SE) is a static program analysis technique that explores program behaviour by executing code over symbolic (rather than concrete) inputs. Although symbolic execution tools and frameworks based on first-order logic, such as CBMC~\cite{cbmc} and KLEE~\cite{klee}, are widely used in academia and industry, their reasoning does not scale when faced with heap-manipulating programs.

\textit{Compositional} symbolic execution~\cite{cse, cse2}~(CSE) addresses this limitation by extending SE with function specifications expressed in a separation logic, such as the original separation logic (SL)~\cite{sl} and the recent incorrectness separation logic (ISL)\cite{isl}. This enables functions to be analysed in isolation, independent of their calling contexts, and records the results as concise specifications for reuse during subsequent analyses. This compositionality significantly improves scalability while retaining precise reasoning about heap-manipulating programs. 

The development and adoption of CSE-based tools has steadily increased both in academia and industry.
For example,
 VeriFast provides semi-automatic compositional verification for C, Java and Rust using SL;
 CN supports verification of C programs using SL; 
 the Viper~\cite{viper} framework, based on implicit dynamic frames (akin to SL), underpins verification tools for Go, Java, Python and Rust~\cite{gobra,nagini,prusti};
 Infer-Pulse~\cite{infer-pulse}, built on ISL, is deployed at Meta for automated multi-language bug finding at scale~\cite{infer-site};
 and Gillian~\cite{gillian1,gillian2,gillian-rust} offers a unified framework for correctness and incorrectness reasoning based on SL and ISL respectively, applied to C, JavaScript and Rust.
These tools have reached a level of maturity that enables their application to real-world software, including the AWS SDK (Gillian), Google's pKVM hypervisor (CN), internet routers (Viper), Meta's production codebase (Infer-Pulse), and the Rust standard library (Gillian, VeriFast).
Several are also used in teaching program analysis principles to undergraduate and graduate students (Gillian, VeriFast, Viper).

With this growing and multi-faceted applicability comes a need for clearly presenting and efficiently debugging the output of a given CSE tool.
To us, this need is threefold:
(1)~to allow tool developers to examine the inner workings of the tool in full detail so that they can improve its implementation;
(2)~to enable tool users to debug their programs by relating the output of the tool to the analysed code; and
(3)~to guide students, who are the next generation of tool users and developers, in understanding the fundamentals of CSE reasoning.
This, however, still remains a relatively uninvestigated topic,  typically addressed in an ad-hoc fashion, with user experience varying greatly from tool to tool:
some only provide (overwhelming) textual feedback; some offer minimal language-server-based integration; others offer interactive exploration but limit it to generated counter-examples or are post-hoc; and the few that do have truly interactive debuggers either adopt broad editor integration at the cost of visual expression or tightly focus on integrating with a single editor, making the transfer to other tools difficult. A more detailed overview of the state-of-the-art is given in \autoref{sec:related-work}.

To address this gap, we introduce an interactive debugger for CSE-based verification, using Gillian as a case study.
Our goal is to design a tool-agnostic debugger, identifying which of its components are general enough to support other CSE tools, and which are specific to Gillian.
Gillian serves as an ideal test bed for this work due to its parametric treatment of memory, support for multiple source languages through compilation to a common intermediate representation (IR), and implementation of multiple analyses within a unified framework~(\autoref{sec:overview}).
Our debugger is integrated with Visual Studio Code~\cite{vscode} to leverage a widely used and familiar development environment, again taking a generalised approach intended to facilitate future integration with additional editors beyond VSCode.

We give a tour of the debugging experience in \autoref{sec:debugger} and present the principles underlying our debugger in \autoref{sec:design}, focusing on:
\begin{itemize}
\item \emph{execution tree representation}, where we aim to create a useful, intuitive, and flexible representation of analysis with a tree-of-trees structure;
\item \emph{execution tree capture}, using this tree-of-trees structure to record information about the analysis in an unobtrusive way at the engine level;
\item \emph{execution tree lifting}, which lifts a low-level tree-of-trees structure to a source-level one, which can be read by the user; and
\item \emph{debugging interactivity}, giving the user full control of examining and directing the symbolic analysis on-the-fly rather than summarising it after completion.
\end{itemize}
Importantly, we take care to decouple our design choices from the Gillian implementation whenever possible, making them transferable to other CSE tools, and even potentially more traditional symbolic execution tools.
Further, in \autoref{sec:impl}, we highlight interesting lessons and observations that have come up during the substantial engineering work involved in this project.

We performed a basic user study at a PL summer school~(\autoref{sec:user-study}), evaluating the usability of the Gillian debugger on CSE-base verification.
Early researchers were introduced to SL, CSE and Gillian, before using Gillian to try to verify a number of data-structure algorithms.
The examples were designed to require use of the main features of semi-automatic compositional verification:
specification of pre- and post-conditions, loop invariants, and proof tactics.
Afterwards, we performed a quantitative analysis that revealed that the users spent a substantial amount of time working with the debugger,
indicating its usefulness.
We also collected qualitative feedback using a Lickert-style questionnaire, with the aggregated results
indicating that working with the debugger was helpful and intuitive, and its feedback informative~and~educational.

%% file: sections/02_related-work.tex

\vspace*{-0.25mm}
\section{Related Work}
\label{sec:related-work}
\vspace*{-0.25mm}

We briefly review the debugging capabilities of state-of-the-art CSE tools.
Gillian (prior to our work) and CN provide only text-based logs, which in real-world verification can reach tens of millions of lines.
These logs linearise the inherently non-linear analysis into flat textual streams that, while exhaustive, are overly detailed and interpretable only by experts familiar with the tools.

Infer-Pulse generates HTML files that present similar information to text logs, organised into analysis ``nodes'' associated with source code locations.
Although this offers greater clarity than plain logs, it remains difficult to follow the flow of the analysis through the program, and the
need to leave the development environment disrupts the user's workflow.

Viper integrates with editors via the Language Server Protocol~(LSP)~\cite{lsp}, re-running analyses as the user types and reporting failures inline.
This proves useful for receiving rapid feedback, though such feedback is often limited to brief error messages.
We argue that this approach is complemented by the availability of a debugger when deeper insight into the analysis is needed.

Finally, VeriFast, through its VSCode extension~\cite{verifast-vscode}, enables interactive inspection of the analysis' execution tree, akin to a debugging session. However, the tree nodes are unlabelled and difficult to relate back to the source code. Moreover, as analysis terminates at the first breakpoint or failure, the post-hoc tree is typically incomplete, leaving some execution branches unexplored.

Our Gillian debugger extends the Debug Adapter Protocol~(DAP)~\cite{dap} to provide an interactive symbolic debugging environment for its CSE analyses.
Several other symbolic analysis tools, such as GobPie~\cite{gobpie-debug} and SecC~\cite{secc-debug}, also use DAP to deliver an integrated debugging experience with many editors ``for free''. However, the DAP offers no intuitive mechanism for representing branching execution, leading ad-hoc solutions such as repurposing the DAP's `threads' view, originally designed for visualising concurrent threads in concrete execution.

The debugger perhaps most similar to ours is the feature-rich Symbolic Execution Debugger~\cite{key-debugger} of the KeY project~\cite{key}. This debugger is capable of displaying detailed, annotated symbolic execution trees, but relies on a non-standard interface and is tightly integrated with both the Eclipse IDE and KeY's first-order Java analyses, limiting its portability to other IDEs and analysis~frameworks.

%% file: sections/03_background.tex

\vspace*{-0.25mm}
\section{Background}
\label{sec:overview}
\vspace*{-0.25mm}

\textbf{Symbolic Execution}
is a program analysis technique that executes a given~program with symbolic instead of concrete inputs, where one symbolic state describes a set of concrete states, enabling general reasoning about program behaviour. Symbolic execution can branch (e.g. via an if-else statement), with each branch carrying a logical assertion called a \emph{path condition} which describes the constraints on symbolic variables that have led the execution to that branch.

\textbf{\textit{Compositional} Symbolic Execution} (CSE) extends symbolic execution to use and create function specifications written in a separation logic.
In particular, the reasoning allows function behaviour to be summarised on \textit{partial} symbolic states using function specifications, which are Hoare triples that have meaning given by an underlying separation logic.
With verification, the pre-condition describes a part of the symbolic state that is sufficient for the function to be evaluated; the post-condition describes all the possible results.

When calling a function, a specification of the callee can be used instead of inlining its body, providing compositionality and improving scalability.
Executing a specification means \emph{matching} the current symbolic state against its pre-condition, \emph{consuming} (or `exhaling') the matched part from the state, and \emph{producing} (or `inhaling') the post-condition.
Verifying a function simply requires symbolically executing the function from its pre-condition until termination, then matching each final state against the post-condition.
Since function calls are replaced by the execution of their specifications, a symbolic execution tree is as long as the number of statements in a single function, drastically reducing reasoning complexity compared to non-compositional SE.

\textbf{Gillian} is a multi-language platform that supports three types of analysis:
 whole-program symbolic testing, providing bounded verification guarantees similarly to the first-order CBMC tool~\cite{cbmc};
 semi-automatic compositional verification à la CN, VeriFast using SL, and Viper;
 and automatic bug-finding similar to Infer-Pulse using ISL.
Gillian is parametric on the memory model of the source language analysed,
which needs to be provided per instantiation together with a compiler from the source language to Gillian's intermediate representation, GIL.
When it comes to real-world languages, Gillian has been instantiated to C, JavaScript, and Rust.
For ease-of-presentation, this paper focuses on \textbf{WISL}, a demonstrator Gillian instantiation for a simple while
language with a C-like block-offset memory model routinely taught to fourth-year undergraduates and~MSc~students.
More detail on this language can be found in \ifExtended{\autoref{apx:wisl} of the appendix}{the extended version of this paper~\cite{arxiv}}.

%% file: sections/04_debug-intro.tex

\vspace*{-0.25mm}
\section{The Gillian Debugging Experience}
\label{sec:debugger}
\vspace*{-0.25mm}

\begin{figure}[!b]
\centering
\vspace*{-0.5cm}
\begin{minipage}{0.5\textwidth}
  \begin{lstlisting}[language=gil, autogobble]
  predicate list(x, alpha) {
    (x == null) * (alpha == nil);
    (x -> #v, #z) * list(#z, #beta)
                  * (alpha == #v::#beta)
  }

  { (x == #x) * list(#x, #alpha) }
  function llen(x) {
    if (x = null) {
      n := 0
    } else {
      t := [x+1]; n := llen(t)
    };
    return n
  }
  { list(#x, #alpha) * (ret == len(#alpha)) }
  \end{lstlisting}
\end{minipage}
\hfill
\begin{minipage}{0.455\textwidth}
\begin{lstlisting}[autogobble]
proc llen(x) {
        goto? (x = null) then else;
  then: n := 0;
        goto end;
  else: _var0 := i_add(x, 1);
        goto? (_var0 is Ptr) cont fail;
  fail: fail "Invalid pointer";
  cont: t := load<_var0>;
        n := llen(t);
  end:  skip;
        ret := n;
        return
};
\end{lstlisting}
\end{minipage}
\vspace*{-0.5cm}
\caption{
  Slightly simplified SLL predicate and a specified WISL recursive list-length function (left), and the corresponding compiled GIL code, simplified (right).
  The prefix \texttt{\#} denotes a logical variable.
}
\label{lst:llen}
\end{figure}

\begin{figure}[!b]
  \centering
  \includegraphics[width=0.95\textwidth]{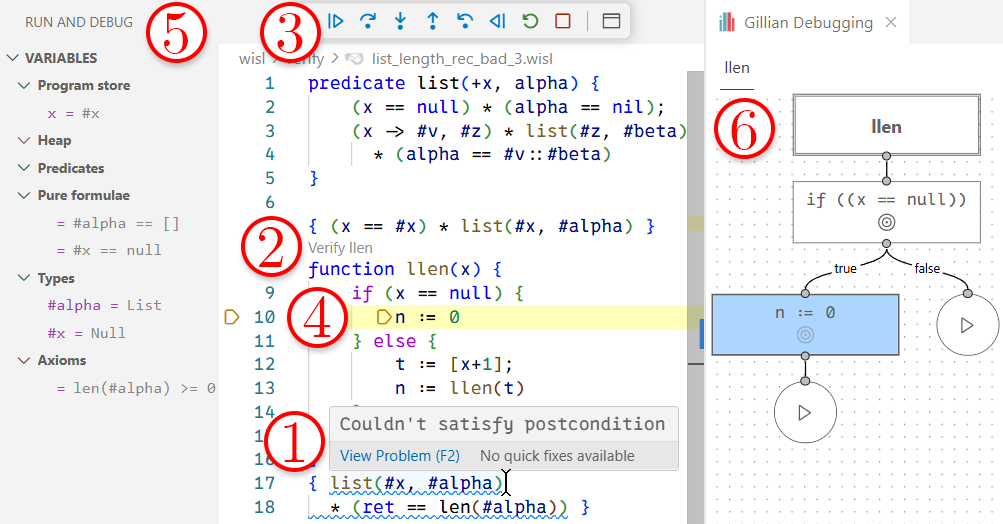}
  \vspace*{-0.2cm}
  \caption{
    Gillian's debugging and language server interface when verifying \texttt{llen()}.
  }
  \label{fig:debug-ui}
\end{figure}

We illustrate the features of the Gillian debugger by using it to debug the verification of the WISL \texttt{llen()} function (\autoref{lst:llen}, left), which computes the length of a null-terminated singly-linked list (SLL), each node of which consisting of a two-cell block containing the value carried by the node and the pointer to the next node.
The function is specified using the standard separation-logic SLL predicate, $\mathtt{list(x, alpha)}$, which states that the SLL starting from block with identifier $\mathtt{x}$ contains the values in the mathematical list $\mathtt{alpha}$.
Then, in the pre-condition we have just the list, and in the post-condition we additionally state that the return value equals its length.
To illustrate the difference between the source- (WISL) and the IR-level (GIL), we also give a stylised compilation of \texttt{llen()} to GIL in \autoref{lst:llen} (right), where we can see how WISL's structured control flow becomes unstructured via GIL gotos, and also how WISL commands (such as pointer dereferencing, $\mathtt{t := [x + 1]}$) get broken down into several~GIL~steps.

When trying to verify \texttt{llen()} using Gillian and VSCode (cf.~\autoref{fig:debug-ui}), the language server informs us that there is an error (\Circled{1}), stating that the function post-condition could not be satisfied. While other existing LSPs, such as Viper's, may provide slightly more precise information, such as which subset of the post-condition failed, the LSP interface can only point to the failure, not explain it.
At this point, we turn to the Gillian debugger to trace the root cause of the error, as the verification could have failed for a variety of reasons, such as an incorrect specification, a missing proof annotation, or a bug in the program itself.

Above each function, the Gillian debugger displays a `Verify' button (\Circled{2}) that initiates an interactive debugging session for that function. While in a session, several pieces of interface are presented to the user: controls for stepping through the executions (\Circled{3}); a highlight of the line of code being currently executed (\Circled{4}); a breakdown of the current symbolic state (\Circled{5}); and a tree view that visualises the symbolic execution presented in a separate panel (\Circled{6}).

The controls at \Circled{3} are inherited from the VSCode DAP interface. They contain: contain: a `continue' button, which continues execution along the current branch until the next breakpoint or end of execution; a `step over' button which executes the next WISL statement; a `step in' button which steps inside a function call if present; a `step out' button which continues execution until the current function returns; a `step back' button which undoes the last statement; a `continue backwards' button which returns to the previous breakpoint or start of execution; a `restart' button which restarts the entire symbolic execution from the beginning; and a `stop' button, which ends the debugging session.

\begin{wrapfigure}{r}{0.37\textwidth}
	\vspace*{-0.8cm}
	\centering
	\includegraphics[scale=0.50]{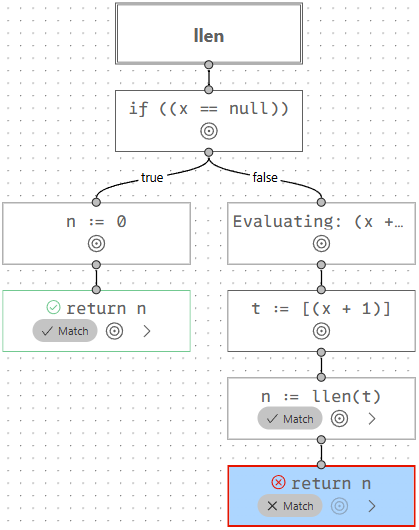}
    \vspace*{-0.6cm}
    \caption{Full symbolic execution tree of \texttt{llen()}}
    \label{fig:debug-ui-trace}
    \vspace*{-0.7cm}
\end{wrapfigure}

When navigating through the symbolic execution using these controls, the line of code currently being executed is highlighted with a small arrow pointing to the sub-expression being evaluated (\Circled{4}). In addition, the current symbolic state is displayed on the left-hand side of the screen, using VSCode's built-in variable explorer to display the current program variable store, heap, predicates, path conditions, and other relevant information (\Circled{5}).

Finally, and most importantly, the tree view (\Circled{6}) enables interactive visualisation of the symbolic execution. Statements are represented as nodes in a tree structure, with edges representing the flow of execution. Clicking on a node jumps to that point in the execution, updating the highlighted code and symbolic state, providing a way to navigate \emph{between} branches as well as forward and backward within the same branch. If a branch has not yet terminated, a `play' button is displayed as a leaf of that branch and can be clicked on to execute the next statement on that branch. When examples increase in complexity and verification times grow longer, this interface allows users to contain branch exploration, avoiding long waiting times.

\autoref{fig:debug-ui-trace} shows the full symbolic execution tree of \texttt{llen()}, from which we identify that the path leading to a verification failure is the one taking the `false' branch of the \texttt{if} statement, as identified by the red cross next to the return statement. This node also displays a small arrow pointing to the right, indicating that there is \emph{nested} information within that node.

\begin{figure}[!b]
  \centering
  \includegraphics[width=0.58\linewidth]{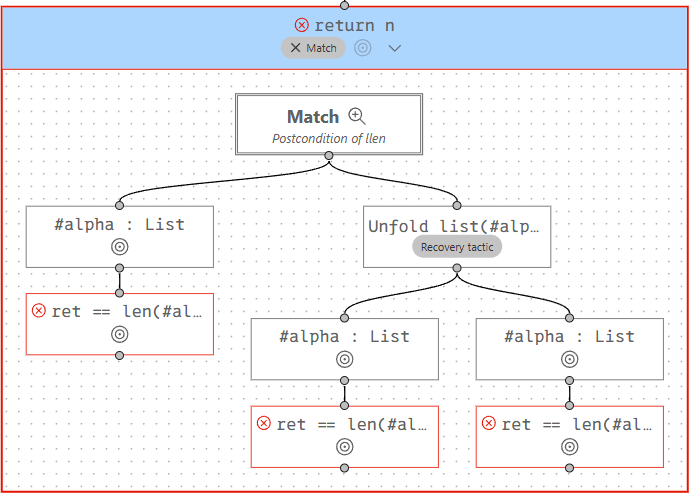}
  \includegraphics[width=0.4\linewidth]{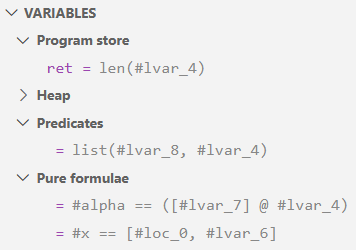}
  \vspace*{-0.2cm}
  \caption{Expanded failing node (left) and symbolic state at the failing match (right).}
  \label{fig:debug-ui-match}
  \vspace*{-0.2cm}
\end{figure}

Clicking that arrow expands the node to reveal the steps of the post-condition matching that failed (\autoref{fig:debug-ui-match}), and the magnifying glass icon in the nested tree `pops it out' to a new tab to view in isolation.
Notice how the visual branching remains relevant in matches, where Gillian has backtracked and attempted to recover by unfolding the list predicate.
Within this tree, we see that Gillian failed to match \texttt{ret == len(\#alpha)} (leftmost branch) before backtracking and applying a \emph{recovery tactic} (here, the `unfold' tactic), where the match still fails for the same reason (two rightmost branches). Recovery tactics are triggered by heuristics and sometimes lead to confusion; this example illustrates how our debugger clearly displays this process to the user.

By clicking on the leftmost failing node and then inspecting the symbolic state in detail (\autoref{fig:debug-ui-match}, right), we see that $\mathtt{ret}$ equals \texttt{len(\#lvar\_4)}, but $\mathtt{\#alpha}$ is the list $\mathtt{([\#lvar\_7] @ \#lvar\_4)}$, and that therefore $\mathtt{ret}$ is off by one.
This is the final step that leads to the understanding of the root cause of the error: we forgot to increment $\mathtt{n}$ in the program itself. In particular, after line 12 of \autoref{lst:llen} (left), $\mathtt{n := n + 1}$ is required.
While this last step must still be taken by the user, the information required to get them there was accessible within a few clicks.

We note that, at the top level, the maximum depth of the execution tree is the number of statements in the function being verified (modulo expression evaluation, which is an implementation choice); as CSE permits verifying functions in isolation, execution tree sizes are constrained. While real-world conditions---e.g. larger functions, more complex source languages---will naturally produce trees larger than we encounter in this work, we expect them to remain orders of magnitude smaller than those seen in non-compositional symbolic analyses \'a la KLEE.

%% file: sections/05_design.tex

\vspace*{-0.25mm}
\section{Principles of CSE Debugging}
\label{sec:design}
\vspace*{-0.25mm}

\begin{figure}[!h]
  \centering
  \vspace*{-0.2cm}
  \includegraphics[width=0.6\textwidth]{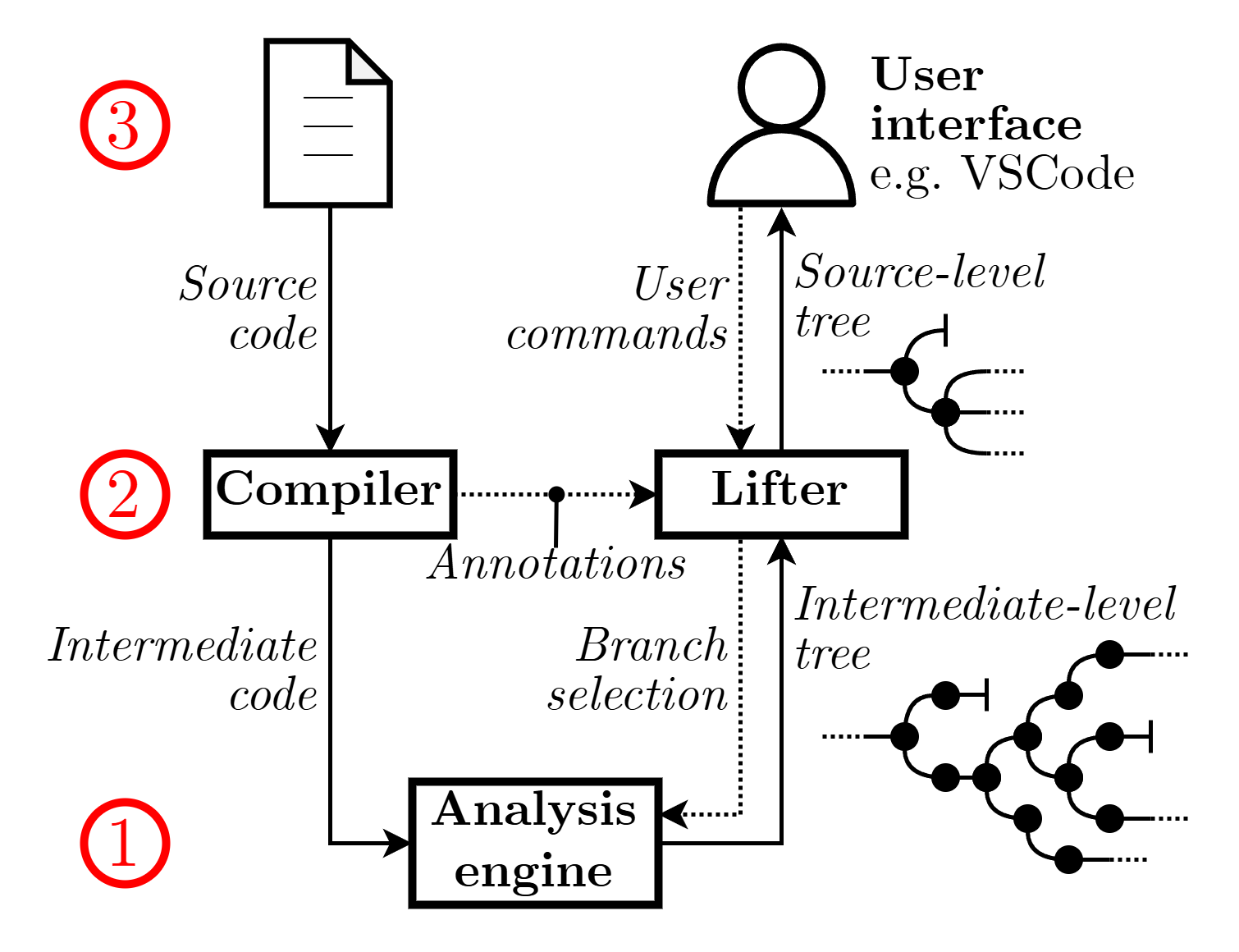}
  \vspace*{-0.3cm}
  \caption{Overview: architecture of the Gillian debugger.}
  \label{fig:overview}
  \vspace*{-0.2cm}
\end{figure}

We give an overview of our three-layered architecture in \autoref{fig:overview} and discuss the key processes we compose to create a usable CSE debugger.
We will refer to the execution of the WISL statements \texttt{t := [x+1]; n := llen(t)} on line 12 of \texttt{llen()}, which correspond to lines 5--8 and 9 of the GIL code respectively (see \autoref{lst:llen}).
We abstract these concepts from Gillian-specific implementation details of Gillian to encourage wider applicability; some implementation details specific to Gillian are discussed in~\autoref{sec:impl}.

At Layer 1, we create a tree structure that describes the analysis process at the level of the intermediate representation (IR).
This is a low-level tree that includes all the execution steps of the analysis, and, ideally, any associated fundamental processes (for Gillian, for example, these would be matching (supported) and expression and state simplifications (currently unsupported)).
This tree must accurately represent one's intuition of the `shape' of the analysis, and be extensible over time in the context of an interactive analysis.
Working with this layer alone is only for the expert tool developer, giving them the ability to fine-tune tool performance and understand unexpected behaviours or bugs in the tool itself.

To present an understandable and actionable execution tree to a tool user working with the analysed source language, one must also bridge the disconnect between the internal IR of the tool and said source language. This is the purpose of Layer 2, at which the debugger employs various lifting mechanisms to abstract the low-level tree from Layer 1 into a source-level execution tree.

Finally, Layer 3 concerns an important but rarely discussed aspect of tool development: the infrastructure through which a user interacts with the tool. UI development, especially when aiming for reuse, maintainability, and longevity, can be an arduous engineering task, as it requires a deep understanding of existing developer tooling.

\subsection{Low-level Tree-of-trees Structure}
\label{sec:design:logs}
An intuitive, interactive debugging experience requires a structure that accurately represents the `shape' of the analysis process, without knowing the execution order ahead of time.
One could argue that symbolic execution can be represented simply as a tree, where a node with multiple children denotes symbolic branching.
We represent analyses at both the intermediate and source level with a \textit{tree-of-trees} structure, where each node can contain nested trees providing further detail.
This has the benefit of capturing the intuition of breaking execution and analysis down into smaller steps, while hiding excessive details until desired by the user.
Nesting is used, for example, to represent Gillian's matching process or the body of an inlined function call; recall \texttt{llen()}'s execution tree given in \autoref{fig:debug-ui-trace}, where the failing \texttt{return} node expanded to reveal the nested matching tree shown in \autoref{fig:debug-ui-match}.

\begin{wrapfigure}{r}{0.35\textwidth}
  \vspace*{-1cm}
  \centering
  \includegraphics[width=0.34\textwidth]{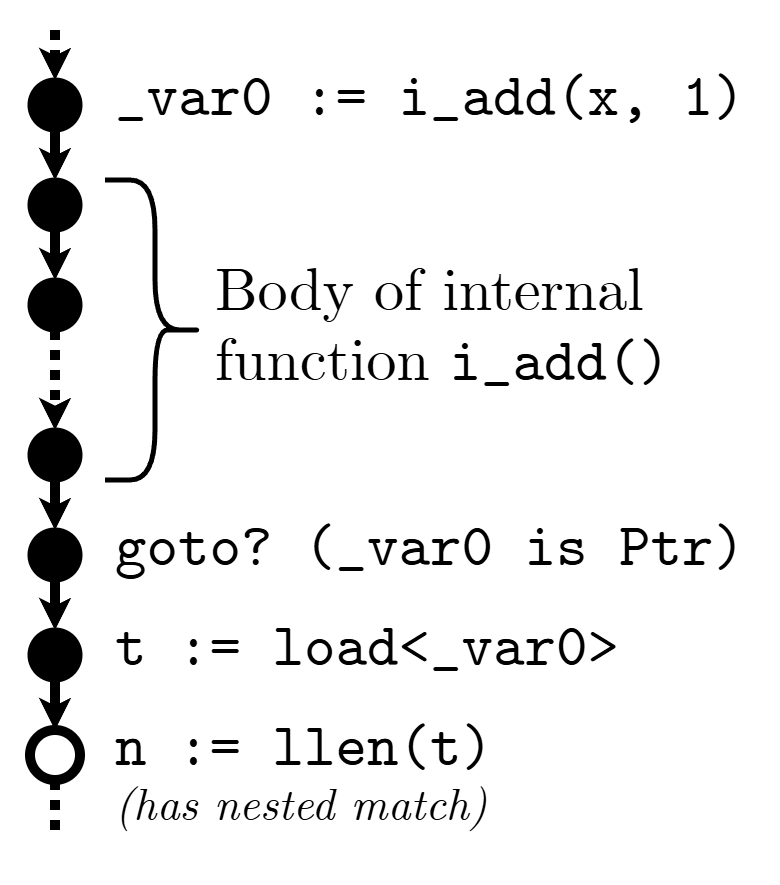}
  \vspace*{-0.55cm}
  \caption{A snippet of \texttt{llen()}'s GIL trace.}
  \label{fig:llen-trace-gil-snippet}
  \vspace*{-0.45cm}
\end{wrapfigure}

A CSE tool will typically implement a particular execution strategy (e.g. depth-first) for traversing the execution tree. Interactive debugging requires recording an execution tree that is agnostic to any such strategy, in such a way that it can handle interrupting and later returning to an execution path.
The tools' analysis engine will therefore need some modification to provide this information necessary to build the tree-of-trees.
However, tracking non-trivial control flow in symbolic execution, such as entering and leaving function and loop bodies, can dramatically increase engine complexity, making a critical part of the tool more difficult to maintain.
We found that a minimal approach is sufficient, and can straightforwardly apply to many CSE tools: instead of just writing text logs describing the state of analysis at each step, like most tools do, the engine can produce a machine-readable report with similar information and a unique identifier.
A report also stores the identifier of the previous execution step, and the enclosing step that nests it (if relevant); this requires only light bookkeeping from the engine and can easily be stored alongside the execution context.
These reports can be freely accessed later in the debugging pipeline, where the heavy lifting of constructing a more intricate tree can be performed with no further input from the engine.
\autoref{fig:llen-trace-gil-snippet} shows a GIL trace segment for the WISL statements on line 12; to keep engine logic simple, function bodies such as that of \texttt{i\_add()} are not nested under the calling command. Nesting is still utilised at the intermediate level where convenient, such as matching for the recursive call to \texttt{llen()}.

\subsection{Lifting to the Source Level}
\label{sec:design:lifting}

After applying the prior concepts, one is rewarded with a perfectly adequate interactive execution tree, but only for the relevant tool's IR.
To facilitate debugging at the level of the source language, a CSE debugger needs to include a \textbf{lifter}, whose responsibility is to tell the analysis engine how to transform the tree-of-trees for a specific IR into a tree-of-trees more reminiscent of a particular source language---in our debugger, from GIL to WISL.
A lifter can, for example, transform the shape of a tree by combining the nodes of multiple IR-level commands into a single node for the originating source-level statement, it could alter and combine branch cases for greater clarity, and it could introduce nesting where appropriate. A lifter can also improve the information associated with individual nodes by, for example, producing node display text and transforming the displayed state to be more recognisable at the source level.

The source-aware information needed to construct a source-level tree-of-trees is most easily gathered during compilation and can be passed to the lifter via \textbf{annotations} on IR code. These annotations may include the display text for the source statement, flags denoting which groups of IR statements comprise a source statement, or hints at the reason a branch may occur.
As an example, the WISL statement \texttt{t := [x+1]} compiles to lines 5--8 of the GIL code: we choose to count expression evaluation as a distinct step, so line 5 is flagged as terminating a node; line 6 is flagged as \textit{not} terminating a node; and lines 7 and 8 are flagged as terminating. In the source-level tree, this results in one step for evaluating \texttt{x+1}, and one for performing the lookup. See \ifExtended{\autoref{apx:wisl-annots} in the appendix}{the extended paper} for some further details on WISL's annotations and their assignments to \texttt{llen()}'s GIL code.

\subsection{User Interface}
\label{sec:design:ui}

When it comes to UI development, given the amount of engineering work it requires, one should try to avoid `reinventing the wheel' to create an interface with a slightly different functionality or in a slightly different environment.

As discussed in \autoref{sec:related-work}, some tools leverage the DAP to take advantage of debugging interfaces provided by IDEs such as VSCode. This includes, for example, highlighting the current code step, controls for stepping through execution, and a view of the current state. The DAP, however, is not expressive enough to represent branching execution, much less navigate a complete execution tree, leading to ad-hoc solutions such as using the `threads' view to list execution branches, or abandoning the DAP altogether in favour of a bespoke post-analysis tree viewer.

Our approach achieves the best of both worlds by \textit{extending} the DAP to support interaction with a custom tree view in the editor.
Referring to \autoref{fig:debug-ui}, the standard debugging UI elements are provided as standard for a DAP-implementing debugger; users can see the line of code (\Circled{4}) and the state (\Circled{5}) at the current step, and use the (linear) stepping controls (\Circled{3}). The bespoke tree view (\Circled{6}) is supported by a custom event \texttt{mapUpdate} to relay changes to the tree, and the custom commands \texttt{jump} to travel to a different point in the tree and \texttt{stepSpecific} to step forward via a specific branch when multiple are available.
This does detract from the DAP's editor agnosticism---due to requiring a VSCode extension---but we believe this to be a worthwhile compromise for the moment, and are working on re-establishing DAP editor independence.

%% file: sections/06_impl.tex

\vspace*{-0.25mm}
\section{Implementation Insights}
\label{sec:impl}
\vspace*{-0.25mm}

While the specifics Gillian Debugging's implementation are not a primary contribution of this work, we would like to highlight a number of observations that have emerged and that could be of interest to the wider community.

\vspace*{-5pt}
\paragraph{Structured logs.}
The structured logs required by the debugger are necessarily more complex to keep track of than regular logs. After experimentation, we decided that the burden of maintaining the structure should be minimised for the symbolic execution engine, an already complex piece of technology with strong correctness requirements. Specifically, we separate concerns: the engine performs minimal bookkeeping, storing all log reports in an SQLite database with only two additional fields, \texttt{previous} and \texttt{parent}; the lifter later uses these reports to construct a tree with more elaborate control flow. The former operation is language-independent and is factored out from Gillian instantiations.

\vspace*{-5pt}
\paragraph{Rebuilding source-level commands.}
The WISL lifter aggregates executed GIL steps into WISL commands.
Each executed GIL step is assigned to a \textit{partial command} representing a source-level WISL command that has not yet been fully explored.
Annotations are used to mark GIL commands that terminate a WISL command, allowing the lifter to finalise a partial command into a source-level tree node when all paths have been explored.
As we had full control over the WISL compiler, we could straightforwardly extend it to add such annotations to the GIL code. This approach is applicable to other compliers: for example, we successfully applied the same pattern to a new Gillian-C instantiation that uses CBMC's parser and IR. We could not adapt the original Gillian-C instantiation, as the CompCert compiler it employs is written in Rocq and is therefore difficult to modify. This demonstrates that our approach can be portable across different language implementations if the frontend preserves sufficient source-level information during compilation.

\vspace*{-5pt}
\paragraph{Advanced nesting.}
The tree-of-trees structure and its visual nesting (cf.~\autoref{fig:debug-ui-match}) were particularly useful in reifying source-level nesting where the IR-level tree had none.
For instance, WISL loops annotated with invariants compile to GIL functions with pre- and post-conditions, but the WISL lifter is able to re-integrate the loop body by initiating its separate verification and nesting the resulting tree in the tree of the outer function.
This demonstrates both the flexibility of the tree-of-trees structure and the ability of the lifter to reconstruct source-level control flow that fundamentally changes during compilation.

\vspace*{-5pt}
\paragraph{Everything is interactive.}
A key difference between our approach and post-analysis tree viewers is the ability for users to interactively guide the analysis as it progresses. To implement this interactivity, we modified Gillian's interpreter to return a continuation function after each GIL execution step. In normal execution mode, these continuations are simply called repeatedly until completion, preserving the original behaviour. In debugging mode, however, the continuations are called with a branch identifier, selecting which execution path to continue.
This approach shows promise for integration with other symbolic execution tools that already operate in continuation-passing style, such as VeriFast and Viper, or with continuation monads, such as CN.

The lifter cooperates with this engine-level interactivity by translating between user-level commands and engine operations. For each high-level instruction (`step in', `step over', etc.), the lifter determines the appropriate GIL steps required to extend the tree.

\vspace*{-5pt}
\paragraph{A generalised UI toolkit.}
Gillian's debugging UI is implemented in VSCode by combining extensions to the DAP with a dedicated tree-viewing interface presented via a web-view panel.
Early iterations of the debugger revealed that both the extensions to the DAP and the custom tree viewing interface had few conceptual dependencies on Gillian or VSCode, and could be separated entirely; the interface deals in abstract steps, requiring no knowledge of Gillian, its IR, or the language being analysed, and said interface is contained in a single web-based UI component.
This realisation has been brought to fruition with the working title of SEDAP (Symbolic Execution DAP), a prototype for a standard extension to the DAP with libraries providing the interactive tree view as an extensible web component, and helpers for integrating this interface with VSCode debug sessions.
Just as the DAP bridges editors with traditional debuggers, further work on SEDAP should bridge editors with analysis tools like those discussed in \autoref{sec:related-work}; Gillian's debugger in VSCode now serves as the first example of this in practice.

\begin{table}[ht]
\centering
\begin{tabular}{L{2.5cm}R{2.2cm}R{1.2cm}R{1cm}R{1.2cm}R{1cm}}
\toprule
\multirow{2}{2.5cm}{\textbf{Benchmark}} & \multicolumn{5}{c}{\textbf{Time to verify (ms)}} \\
& \textbf{No logs} & \multicolumn{2}{c}{\textbf{File}} & \multicolumn{2}{c}{\textbf{Database}} \\
\midrule
\texttt{SLL\_iterative.wisl} & 242  & 680   & (2.8x)  & 487  & (2.0x)  \\
\texttt{DLL\_recursive.wisl} & 91   & 259   & (2.8x)  & 174  & (1.9x)  \\
\texttt{sll.c}               & 303  & 390   & (12.8x) & 668  & (2.2x)  \\
\texttt{dll.c}               & 1875 & 81501 & (43.5x) & 6335 & (3.4x)  \\
\texttt{priQ.c}              & 292  & 5730  & (19.6x) & 7815 & (2.7x)  \\
\texttt{sort.c}              & 236  & 2369  & (10.0x) & 475  & (2.0x)  \\
\texttt{SLL.js}              & 149  & 1192  & (8.0x)  & 1188 & (8.0x)  \\
\texttt{DLL.js}              & 202  & 1624  & (8.1x)  & 1672 & (8.3x)  \\
\texttt{ExprEval.js}         & 246  & 6415  & (26.1x) & 4513 & (18.3x) \\
\bottomrule
\end{tabular}
\vspace{3pt}
\caption{
  Benchmark results on a selection of Gillian's verification examples across WISL, Gillian-C and Gillian-JS.
  This compares no logging, verbose file logging, and structured database logging, showing slowdown factor for the latter two.
  Benchmarks were performed on a laptop with an Intel i7-1065G7 processor and 32 GiB of memory.
}
\label{tbl:benchmarks}
\vspace*{-0.6cm}
\end{table}

\vspace*{-5pt}
\paragraph{Performance impact.}
As shown in \autoref{tbl:benchmarks}, structured logging to the on-disk database incurs a performance cost over no logging, but is no worse than that of Gillian's default text log.
Interestingly, database logs fare much better against file logs in C verification, owing to the complicated logic used to print the heap; this demonstrates a benefit of separating logging concerns from the engine.
We note that no effort has yet been made to optimise structured logs, and there is much low-hanging fruit for performance improvements: database queries could be performed asynchronously or in batches; a different database model could be more appropriate (say, a graph database rather than relational); and a more efficient serialisation method such as protocol buffers could be used instead of~JSON.

%% file: sections/07_user-study.tex

\vspace*{-0.25mm}
\section{Empirical Evaluation}
\label{sec:user-study}
\vspace*{-0.25mm}

We made a formative attempt at evaluating Gillian debugging as part of a summer school attended by late-stage undergraduate and early-stage PhD students interested in formal methods and PL research.
Following 3 hours of lectures on SL, CSE and Gillian, participants were invited to spend two 90-minute lab sessions during which they would attempt to verify a number of WISL programs using Gillian's language server and debugger.
For this lab, we created a suite of 13 introductory and 6 advanced exercises: the introductory ones use simple programs on linked lists to introduce various aspects of Gillian verification, whereas the advanced ones consist of larger programs using more complex data structures (e.g., sets, doubly-linked-lists, and binary search trees) as well as a WISL translation of a part of the Collections-C library~\cite{collections-c}; see \ifExtended{\autoref{apx:exercises} in the appendix}{the extended paper} for the full list of exercises.

The lab session was initially guided in that we solved the first few exercises together with the participants, introducing them to the debugger interface and the Gillian proof tactics that they had at their disposal: folding and unfolding user-defined predicates, the \texttt{assert} and \texttt{apply} proof tactics, and loop invariants.

At the end of proceedings, participants were invited to complete an anonymised survey about their experience and to optionally submit usage logs.
These logs were generated and stored locally on the participants' devices and tracked a small amount of timestamped information per participant interaction with the debugger and language server.
Of the $\approx$30 lab participants, 19 completed the survey, of which 18 submitted usage logs.
We use this data to evaluate the usefulness of Gillian debugging in understanding CSE and completing the provided exercises.
We first evaluate qualitatively by analysing responses to a series of Lickert-scale statements about the lab to gauge subjective participant impressions on the usefulness of the debugger.
We then contrast these responses against a quantitative metric by analysing the provided usage logs to understand the actual participant use of the debugger and language server.

The provided Lickert-style statements and their combined responses are shown in \autoref{tbl:lickert}.
The overall sentiment towards the usefulness of the debugger was overwhelmingly positive (statements 1/5/6), with 63\% of responses strongly in favour and 26\% somewhat so.
Further, the participants felt that the debugger was well-integrated with the verification process (statement 10) and that other tools could benefit from similar debuggers (statements 11/12).
The participants also recognised the usefulness of more specific properties arising from the interactive, step-by-step nature of the debugger, including the exploration of individual execution paths (statement~7) and viewing the state at each execution step (statement~9).
They were less positive, however, about the viability of the language server alone (statements 2/8), with neutral-to-slightly-negative~responses.

\begin{table}[!t]
\centering
\newcommand{\rbox}[1]{\rotatebox[origin=c]{90}{\parbox{1.5cm}{\textbf{#1}}}}
\renewcommand{\arraystretch}{0}
\renewcommand{\baselinestretch}{0.9}\selectfont
\begin{tabular}{R{0.5cm}L{7cm}R{0.8cm}R{0.8cm}R{0.8cm}R{0.8cm}R{0.8cm}}
 & & \rbox{Strongly\\Disagree} & \rbox{Somewhat\\Disagree} & \rbox{Neutral} & \rbox{Somewhat\\Agree} & \rbox{Strongly\\Agree} \\
\noalign{\vskip 1mm}
\toprule
1\; & I didn't need to use the debugger to complete the exercises & 11 & 6 & 1 & 1 & 0 \\
\midrule
2\; & The language server was sufficient for me to complete the exercises & 1 & 6 & 5 & 6 & 1 \\
\midrule
3\; & The debugger helped me to understand CSE & 0 & 2 & 5 & 5 & 7 \\
\midrule
4\; & The debugger helped me understand what Gillian is doing & 0 & 0 & 2 & 8 & 9 \\
\midrule
5\; & I found the debugger helpful in completing the exercises & 0 & 0 & 3 & 4 & 12 \\
\midrule
6\; & The debugger helped me to understand why verification failed & 0 & 2 & 3 & 6 & 8 \\
\midrule
7\; & I found it useful to explore each path of execution individually using the debugger & 0 & 0 & 1 & 5 & 13 \\
\midrule
8\; & I understood the errors that the language server presented me with & 1 & 7 & 3 & 4 & 4 \\
\midrule
9\; & I found it useful to see the state at each point in execution using the debugger & 0 & 0 & 3 & 5 & 11 \\
\midrule
10\; & It felt natural to use the debugger as part of the verification workflow & 0 & 1 & 2 & 4 & 12 \\
\midrule
11\; & This debugging interface would be applicable to similar tools I have used & 0 & 1 & 4 & 3 & 6 \\
\midrule
12\; & I would benefit from a debugger like this in similar tools I have used & 0 & 0 & 1 & 4 & 9 \\
\bottomrule
\end{tabular}
\vspace{3pt}
\caption{Summed Lickert scale responses of the user survey.}
\label{tbl:lickert}
\vspace*{-0.5cm}
\end{table}

On the quantitative side, the usage logs consist of timestamped entries for each participant interaction with the debugger and the language server; language server entries show the changes to the exercise file and the analysis result (success, or the reason for a failure), and debugger entries denote whether the user started or stopped a session or made a step through the tree.
From this, we calculate the amount of time each participant spent on each question (accounting for switching between exercises and long periods of inactivity), and how much of that time was spent actively using the debugger (interpreted as having interacted with the debugger within the last 30 seconds), to build a rough estimate of what proportion of time participants spent using the debugger.
The distribution of these proportions, shown in \autoref{fig:debugger-usage-hist}, suggests that the majority of participants spent between 40\% and 60\% of their time having recently interacted with the debugger, giving a strong indication that the participants indeed made frequent and active use of the debugger.

\begin{figure}[!t]
  \centering
  \includegraphics[width=0.7\textwidth]{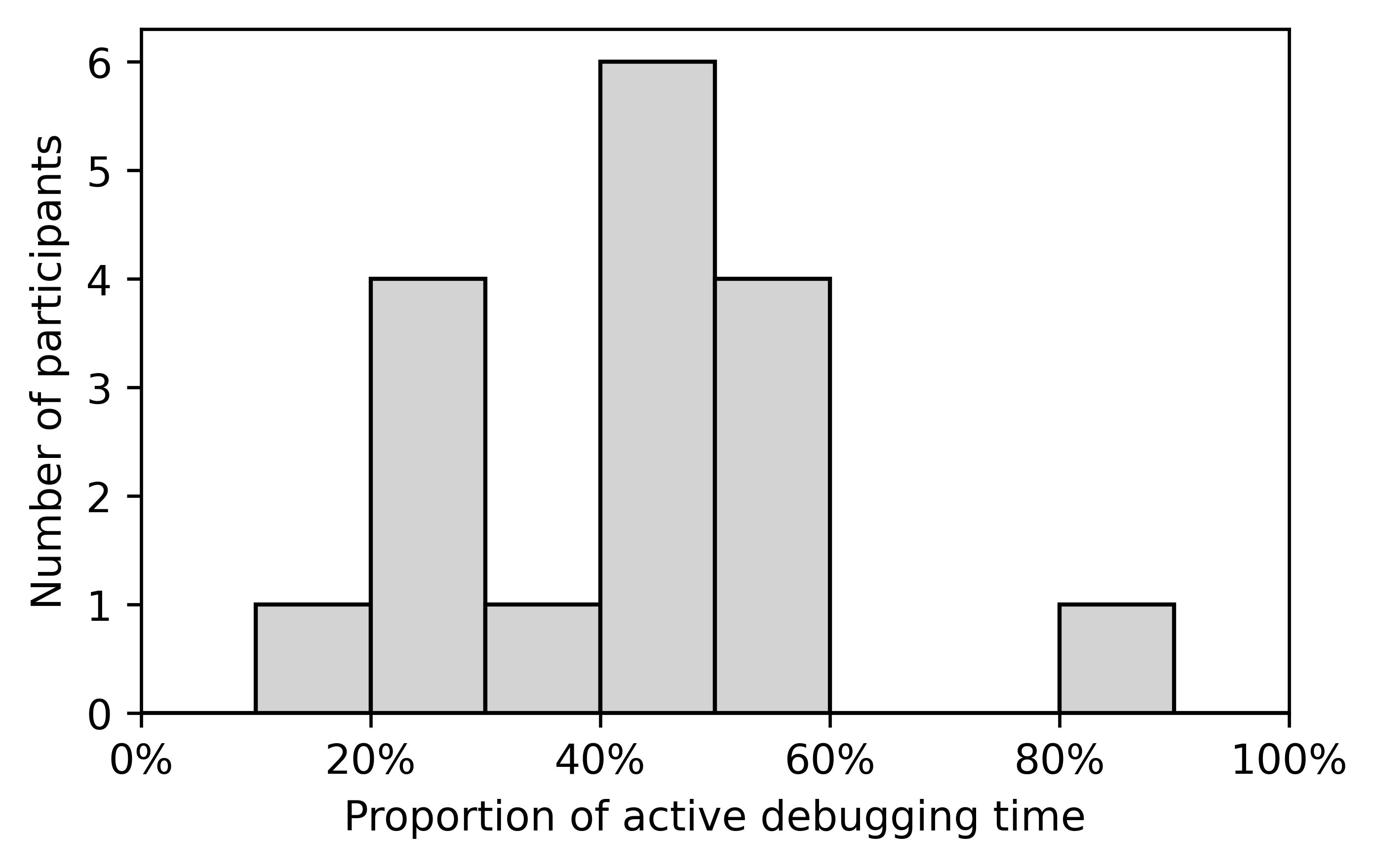}\vspace*{-0.3cm}
  \caption{
    Distribution: proportion of participant time in an active debugging session.
  }
  \label{fig:debugger-usage-hist}
  \vspace*{-0.3cm}
\end{figure}

While this user study provided solid evaluation for Gillian debugging, a number of limitations prevented deeper insight.
Ultimately, the lab primarily aimed to be engaging and informative for the participants, meaning the controlled conditions and intense observation desirable of a more formal user study were not feasible. Participants were not required to focus entirely on the exercises at hand, nor work individually, nor stay for the whole runtime. Collecting more detailed information about participant behaviour would be difficult without more invasive techniques such as screen recording and diaries, which are arguably inappropriate in this context.
Additionally, the students were still at an early stage in their PL-specific eduction; at least 60\% of participants reported no prior experience with each of proof assistants, symbolic execution --based tools, CSE --based tools, and separation logic (see \ifExtended{\autoref{apx:user-experience} in the appendix}{the extended paper}). In contrast, the suite of exercises was created based on prior demonstrations of Gillian debugging to attendees of a fourth-year undergraduate course on separation logic. This resulted in less progress through the exercises than desired, with participants completing up to 6 of the 13 introductory exercises, and not reaching the advanced exercises.

After discussion with students and analysing these results, we believe that Gillian debugging has greater potential than simply assisting with verification tasks.
We have formed new hypotheses on its use as a powerful tool for learning about CSE and its practical applications.
For instance, we believe that newcomers to Gillian would use the debugger much more extensively than advanced users, who would reserve its use for more complex debugging tasks.
While this hypothesis seems consistent with the positive responses to statements 3, 4 and 6, further study is required to assert it with greater confidence.
Similarly, the participants' inexperience with other CSE tools, together with the responses to statements 11 and 12, suggest that interactive visualisation could prove helpful for other analysis techniques such as theorem proving or abstract interpretation.

%% file: sections/08_future.tex

\vspace*{-0.25mm}
\section{Lessons Learned \& Future Work}
\label{sec:future}
\vspace*{-0.25mm}

Our experience developing and evaluating the symbolic execution debugger has revealed potential in making CSE tools more accessible, while highlighting work that needs to be done.
We reflect on the lessons learned through this project, and outline directions for future research.

\vspace*{-6pt}
\paragraph{User study.}
Performing a user study is a rare occurrence in this space, and the efforts to perform one in this work were perhaps unexpectedly valuable; despite an evaluation of Gillian debugging being its primary motivation, it resulted in interesting discussions on the nature of tool usability and how it could vary for different analyses and levels of user experience.
In addition to demonstrating educational value, future studies might target experienced tool users performing real-world verification tasks, observing in finer detail to perform focused evaluations of specific design choices.
It may also be prudent to observe more objective metrics, such as the overall size or amount of branching in resulting trees.

\vspace*{-6pt}
\paragraph{Gillian.}
Due to its highly modular nature, Gillian proved to be a useful test bed for exploring debugging in this style, naturally leading to the separation of responsibilities between the compiler, analysis engine, and lifter. Its language-agnosticism provides the opportunity to explore the creation of symbolic execution trees for other languages without re-implementing the debugger internals, particularly to investigate debugging analyses of real-world programs using Gillian’s implementations for JavaScript and C. To this effect, a prototype lifter for our second Gillian-C has been created with the same techniques used for WISL. In addition to semi-automatic verification, Gillian can perform automatic, bounded, whole-program symbolic testing (with non-compositional symbolic execution) and true bug-finding with bi-abduction; further work will be able to explore the applicability of the debugging interface introduced in this work to these analyses.

\vspace*{-6pt}
\paragraph{Scaling analyses.}
Our focus on small, educational verification examples naturally raises the question of scaling to real-world programs, both in terms of lifter complexity for a real-world language and the interface's ability to elegantly handle larger execution trees.
Future work will apply our approach to analyses on non-trivial C programs, via both Gillian and other analysis tools such as CN.

\vspace*{-6pt}
\paragraph{Presenting the symbolic state.}
While this work primarily focuses on presenting an intuitive shape of analysis, more attention will be needed on the presentation of the symbolic state. For example, the WISL lifter prints the GIL-level expressions produced by the engine in a more WISL-esque syntax, but the state still includes intermediate variables introduced during compilation. These variables can be confusing to users as they seemingly have no basis in the source program, but are a necessary part of Gillian’s analysis on its intermediate representation. An investigation into how to meaningfully display CSE states at the source level could prove useful in improving user experience.

\vspace*{-6pt}
\paragraph{Error messages.}
Error messages are a notable pain point of this implementation, corroborated by the user survey (see \autoref{tbl:lickert}, statement 8).
Verification error messages are difficult to implement, as it is often impossible to differentiate between an erroneous implementation, an erroneous specification, or insufficient proof annotations. Nonetheless, we do believe that Gillian's error messages have room for improvement.
Interactive debugging could turn an immutable error message into a conversation with the tool, exploring the entire analysis tree to refine the error into something more specific.

\vspace*{-6pt}
\paragraph{UI generality.}
Our DAP extensions and custom UI have already been decoupled from Gillian and VSCode to form the SEDAP.
Further work will realise the goal of SEDAP to provide debugging for many symbolic analysis tools, connected to many well-known editors.
Our hope is that this will in turn expand SEDAP's capability to represent different internal processes, types of analyses, or volumes of information.
How would SEDAP fare with abstract interpretation? Could it be extended to display symbolic heap visualisations? Could interactivity be leveraged to, for example, apply proof tactics in a live debugging session?

%% file: sections/09_conclusion.tex
\vspace*{-0.25mm}
\section{Conclusion}
\label{sec:conclusion}
\vspace*{-0.25mm}

We have introduced a novel debugging interface for Gillian, addressing the underexplored challenge of creating intuitive debugging interfaces for compositional symbolic execution tools.
We focus on providing a familiar user experience by leveraging VSCode's existing debugging interface for standard UI elements, while developing custom components specifically designed to present compositional symbolic execution information in an intuitive way that accurately captures the analysis flow, including branching, nested information and state matching.

We have given the conceptual principles underlying our design, emphasising transferability of the approach to other tools, and discussed interesting technical insights that arose during implementation in the hope that they may inform future endeavours in this space. Importantly, we have decoupled the UI from Gillian and VSCode as much as possible, resulting in an independent library which we hope to continue developing and see adopted by other tools and editors.

Finally, we have conducted a preliminary evaluation of the debugger through a quantitative and qualitative analysis of its use by early researchers during a lab session, providing validation of its usefulness in understanding and working with compositional symbolic execution.

While the amount of engineering work involved to start proper research on improving the usability of compositional symbolic execution tools is significant---and several attempts have been made in the past and later abandoned---we hope that this work will serve as a stepping stone for future research in this space.

%% file: sections/appendix.tex
\vspace*{-0.25mm}
\section{WISL}
\label{apx:wisl}
\vspace*{-0.25mm}

\subsection{Syntax}
\noindent\textbf{Values}
$$v \in \vals_{W}~\eqdef~\nil \mid n \in \Nat \mid b \in \Bool \mid \ptr \in \wlocs \times \Nat$$

\noindent\textbf{Expressions}
\[
\begin{array}{lll}
E \in \exprs_W  & \eqdef & v \in \vals_{W} \mid \pvar x \in \PVar \mid \unop E \mid E \binop E 
\end{array}
\]

\noindent\textbf{Commands}
\[
\begin{array}{l@{~}l@{~}ll}
\scmd \in \cmds{W} & \eqdef & 
	\pskip  \mid
	\passign{\pvar{x}}{\expr{E}} \mid
            \pfuncall{\pvar{x}}{f}{\vec{\expr{E}}} & \text{Basic} \\[.1cm]
            & &
            \mid \pderef{\pvar{x}}{\expr{E}}\mid
            \pmutate{\expr{E}}{\expr{E}} \mid
            \palloc{\pvar{x}}{\expr{E}} \mid
            \pdealloc{\expr{E}} & \text{Memory management}\\[.1cm]
&&           \mid \scmd; \scmd\mid
             \pifelse{E}{\scmd}{\scmd}\mid
             \pwhile{E}{\scmd} & \text{Control flow} 
\end{array}
\]

\noindent\textbf{Functions}
$$\procedure{f}{\vec{\expr{E}}}{\scmd; \preturn{\expr{E}}} \in \procs{W}$$

\subsection{Annotations}
\label{apx:wisl-annots}

\begin{minted}[fontsize=\scriptsize]{ocaml}
  (** How does this command map to a WISL statement? *)
  type stmt_kind =
    | Normal of bool
        (** A command that makes up part of a WISL statement, and whether this
            command terminates said statement *)
    | Return of bool
        (** Same as Normal, but specific to a WISL return statement *)
    | Hidden  (** A command that doesn't map to a particular WISL statement *)
    | LoopPrefix  (** Special case for initial commands of a loop body function *)

  (** What kind of branch can occur at this command? *)
  type branch_kind =
    | IfElse
    | WhileLoop

  (** What kind of nesting can occur at this command? *)
  type nest_kind =
    | LoopBody of string
        (** Special case for WISL's loop body abstraction *)
    | FunCall of string
        (** Source-level function call; used to inform nesting when calling a
            function non-compositionally *)

  type annot = {
    source_loc : Location.t option;
    stmt_kind : stmt_kind;
    branch_kind : branch_kind option;
    nest_kind : nest_kind option;
  }
\end{minted}

\begin{table}[H]
\centering
\setlength{\tabcolsep}{5pt}
\begin{tabular}{rlll}
\toprule
\textbf{Line} & \textbf{Statement} & \textbf{\texttt{stmt\_kind}} & \textbf{\texttt{branch\_kind}} \\
\midrule
 2 & \texttt{goto? (x = null)} \ldots      & \texttt{Normal true}  & \texttt{Some IfElse} \\
 3 & \texttt{n := 0}                       & \texttt{Normal true}  & \texttt{None} \\
 4 & \texttt{goto end}                     & \texttt{Hidden}       & \texttt{None} \\
 5 & \texttt{\_var0 := i\_add(x, 1)}       & \texttt{Normal true}  & \texttt{None} \\
 6 & \texttt{goto? (\_var0 is Ptr)} \ldots & \texttt{Normal false} & \texttt{None} \\
 7 & \texttt{fail "Invalid pointer"}       & \texttt{Normal true}  & \texttt{None} \\
 8 & \texttt{t := load<\_var0>}            & \texttt{Normal true}  & \texttt{None} \\
 9 & \texttt{n := llen(t)}                 & \texttt{Normal true}  & \texttt{None} \\
10 & \texttt{skip}                         & \texttt{Hidden}       & \texttt{None} \\
11 & \texttt{ret := n}                     & \texttt{Return false} & \texttt{None} \\
12 & \texttt{return}                       & \texttt{Return true}  & \texttt{None} \\
\bottomrule
\end{tabular}
\vspace{3pt}
\caption{Annotations on each GIL command of \texttt{llen()} (see \autoref{lst:llen}).}
\label{apx:llen-annots}
\end{table}

\vspace*{-0.25mm}
\section{Further Evaluation Details}
\vspace*{-0.25mm}

\begin{table}[H]
\centering
\begin{tabular}{L{0.4\textwidth}R{0.12\textwidth}R{0.12\textwidth}R{0.12\textwidth}R{0.12\textwidth}}
\toprule
\textbf{Area} & \textbf{None} & \textbf{<1yr} & \textbf{1--3yrs} & \textbf{>3yrs} \\
\midrule
Proof assistants & 12 & 5 & 1 & 1 \\
Symbolic Execution –based tools & 13 & 5 & 1 & 0 \\
CSE –based tools & 18 & 1 & 0 & 0 \\
Separation Logic & 16 & 3 & 0 & 0 \\
\bottomrule
\end{tabular}
\vspace{3pt}
\caption{The amount of prior experience in related areas reported by lab participants.}
\label{apx:user-experience}
\end{table}

\subsection{Lab Exercises}
\label{apx:exercises}

Here, we list the exercises created for the lab featured in our evaluation, with a count of how many participants reported that they attempted and/or completed each.

\begin{enumerate}
\item{
  \textbf{Intro (auto)} ---  19 attempted, 19 completed \\
  Recursive list length, with a small mistake; similar to \autoref{lst:llen}.
  This example was briefly demonstrated live, and then provided as a trivial example for participants to experiment and familiarise themselves with Gillian and the debugger.
}
\item{
  \textbf{Intro (manual)} --- 19 attempted, 19 completed \\
  Recursive list length, now corrected, but with manual mode enabled.
  This was also demostrated live; it introduces proof tactics, and demonstrates how Gillian handles predicates via folding and unfolding.
}
\item{
  \textbf{Folding} --- 16 attempted, 2 completed \\
  An interesting edge case of manual folding, where the base case of a linked list must be manually folded, even though it doesn't consume anything from the state.
}
\item{
  \textbf{Function calls} --- 15 attempted, 1 completed \\
  An example of using fold/unfold to inductively build a predicate using the result of a function call.
}
\item{
  \textbf{Branching \& recursion} --- 9 attempted, 4 completed \\
  Demonstrates the need to apply different proof tactics in different branches.
}
\item{
  \textbf{Assert} --- 0 attempted, 0 completed \\
  Introduces the \texttt{assert} tactic, used to ensure a condition holds, and to bind logical variables for later reference.
}
\item{
  \textbf{Apply} --- 0 attempted, 0 completed \\
  Introduces lemmas, and how they are used with the \texttt{apply} tactic.
}
\item{
  \textbf{List-remove} --- 0 attempted, 0 completed \\
  The first of a group of exercises on \texttt{list\_remove()}. Tests understanding of fold \& unfold.
}
\item{
  \textbf{Auto mode} --- 0 attempted, 0 completed \\
  Follows on from the previous exercise.
  One is asked to paste their previous solution, and then see which tactics can be omitted now that manual mode is disabled.
}
\item{
  \textbf{Stronger spec} --- 0 attempted, 0 completed \\
  Follows on from the previous exercise.
  It is pointed out that the spec of \texttt{list\_remove()} is weak, and one is tasked with writing a predicate used to strengthen the spec.
}
\item{
  \textbf{Spec change} --- 0 attempted, 0 completed \\
  Follows on from the previous exercise.
  A desired change to the behaviour of \texttt{list\_remove()} is described; one must make the required changes to the function body, and update the post-condition accordingly.
}
\item{
  \textbf{Invariants} --- 0 attempted, 0 completed \\
  An introduction to loop invariants in WISL. The loop invariants are provided, and participants must insert tactics to ensure the proofs both inside and outside the loop body succeed.
}
\item{
  \textbf{Fib} --- 0 attempted, 0 completed \\
  The last of the introductory exercises.
  Uses an iterative fibonacci function to further familiarise with loop invariants, while demonstrating that `pure' predicates (and lemmas that fold them) can be duplicated.
}
\item{
  \textbf{List to set} --- 0 attempted, 0 completed \\
  The first of the advanced exercises.
  Concerns the creation of a `set-list': a list with no duplicate elements that can function as a set, with a predicate describing this property.
  A bonus challenge asks one to find a weakness of the given specification, and then strengthen it.
}
\item{
  \textbf{Doubly-linked lists} --- 0 attempted, 0 completed \\
  Verifies \texttt{concat()} for a doubly-linked list; involves complicated lemma applications.
}
\item{
  \textbf{Well-bracketed} --- 0 attempted, 0 completed \\
  Given a function that checks the well-bracketed-ness of a string, one must write a predicate describing the well-bracketed property.
}
\item{
  \textbf{Binary search trees} --- 0 attempted, 0 completed \\
  Verifies a number of functions on binary search trees, requiring substantial assertions and lemma applications.
}
\item{
  \textbf{Collections-C SLL} --- 0 attempted, 0 completed \\
  Gives a taste of verifying real world --style code with a WISL translation of some singly-linked list functions from Collections-C~\cite{collections-c}.
}
\item{
  \textbf{More loop invariants} --- 0 attempted, 0 completed \\
  Presents some of the earlier singly-linked-list algorithms iteratively instead of recursively.
}
\end{enumerate}